\documentclass[twocolumn]{aastex631}

\usepackage[utf8]{inputenc} 
\usepackage[T1]{fontenc}    
\usepackage{ulem}
\usepackage{amsmath}  
\usepackage{tabularx} 
\usepackage{physics} 

\usepackage{booktabs}

\begin{document}

\title{Searching for pulsars in Globular Clusters with the Fast Fold Algorithm and a new pulsar discovered in M13}

\author[0009-0007-6770-5899]{Yaowei Li}
\affiliation{College of Physics, Guizhou
University, Guiyang 550025, P.R. China}

\author[0000-0003-0757-3584]{Lin Wang}
\affiliation{Shanghai Astronomical Observatory, Chinese Academy of Sciences, 80 Nandan Road, Shanghai 200030, P.R. China}
\affiliation{Kavli Institute for Astronomy and Astrophysics, Peking University, Beijing 100871, P.R. China}

\author[0000-0003-0597-0957]{Lei Qian}
\affiliation{Guizhou Radio Astronomical Observatory, Guizhou University Guiyang 550025, P.R. China}
\affiliation{National Astronomical Observatories, Chinese Academy of Sciences, 20A Datun Road, Chaoyang District, Beijing 100101, P.R. China}
\affiliation{CAS Key Laboratory of FAST, National Astronomical Observatories, Chinese Academy of Sciences, Beijing 100101, P.R. China}
\affiliation{College of Astronomy and Space Sciences, University of Chinese Academy of Sciences, Beijing 100049, P.R. China}

\author[0000-0002-2394-9521]{Liyun Zhang}
\affiliation{College of Physics, Guizhou University, Guiyang 550025, P.R. China}
\affiliation{International Centre of Supernovae, Yunnan Key Laboratory, Kunming 650216, P.R. China}

\author{Yujie Chen}
\affiliation{College of Physics, Guizhou University, Guiyang 550025, P.R. China}

\author[0000-0001-6051-3420]{Dejiang Yin}
\affiliation{College of Physics, Guizhou
University, Guiyang 550025, P.R. China}

\author[0009-0008-4109-744X]{Baoda Li}
\affiliation{College of Physics, Guizhou University, Guiyang 550025, P.R. China}

\author[0009-0007-6396-7891]{Yinfeng Dai}
\affiliation{College of Physics, Guizhou University, Guiyang 550025, P.R. China}

\author[0000-0001-6196-4135]{Ralph~P.~Eatough}
\affiliation{National Astronomical Observatories, Chinese Academy of Sciences, 20A Datun Road, Chaoyang District, Beijing 100101, P.R. China}
\affiliation{Max-Planck-Institut f\"{u}r Radioastronomie, Auf dem H\"{u}gel 69, D-53121, Bonn, Germany}

\author{Wenze Li}
\affiliation{College of Physics, Guizhou University, Guiyang 550025, P.R. China}

\author{Dongyue Jiang}
\affiliation{College of Physics, Guizhou University, Guiyang 550025, P.R. China}

\author{Xingnan Zhang}
\affiliation{State Key Laboratory of Public Big Data, Guizhou University, Guiyang 550025, P.R. China}

\author{Minghui Li}
\affiliation{State Key Laboratory of Public Big Data, Guizhou University, Guiyang 550025, P.R. China}

\author[0009-0001-6693-7555]{Yujie Lian} 
\affiliation{Institute for Frontiers in Astronomy and Astrophysics, Beijing Normal University, Beijing 102206, P.R. China}
\affiliation{School of Physics and Astronomy, Beijing Normal University, Beijing 100875, P.R. China}

\author{Yuxiao Wu}
\affiliation{Chongqing University of Posts and Telecommunications, Chongqing, 40000, P.R. China}

\author{Tong Liu}
\affiliation{National Astronomical Observatories, Chinese Academy of Sciences, 20A Datun Road, Chaoyang District, Beijing 100101, P.R. China}

\author[0000-0002-2953-7376]{Kuo Liu}
\affiliation{Shanghai Astronomical Observatory, Chinese Academy of Sciences, 80 Nandan Road, Shanghai 200030, P.R. China}
\affiliation{100101 Key Laboratory of Radio Astronomy and Technology (Chinese Academy of Sciences), A20 Datun Road, Chaoyang District, Beijing, 100101, P. R. China}

\author[0000-0001-7771-2864]{Zhichen Pan}
\affiliation{Guizhou Radio Astronomical Observatory, Guizhou University Guiyang 550025, P.R. China}
\affiliation{National Astronomical Observatories, Chinese Academy of Sciences, 20A Datun Road, Chaoyang District, Beijing 100101, P.R. China}
\affiliation{CAS Key Laboratory of FAST, National Astronomical Observatories, Chinese Academy of Sciences, Beijing 100101, P.R. China}
\affiliation{College of Astronomy and Space Sciences, University of Chinese Academy of Sciences, Beijing 100049, P.R. China}

\correspondingauthor{Lin Wang, Lei Qian, Liyun Zhang}
\email{wanglin@shao.ac.cn; lqian@nao.cas.cn; liy\_zhang@hotmail.com}

\begin{abstract}
We employed the Fast Folding Algorithm (FFA) on L-Band Globular Cluster (GC) observations taken with Five-hundred-meter Aperture Spherical radio Telescope (FAST) to search for new pulsars, especially  those with a long rotational period. We conducted a search across 16 GCs that collectively host 93 known pulsars, as well as 14 GCs that do not contain any known pulsars. The majority of these known pulsars were successfully re-detected in our survey. The few non-detections could be attributed to the high accelerations of these pulsars. Additionally, we have discovered a new binary millisecond pulsar, namely M13I (or PSR J1641+3627I) in GC M13 (or NGC 6205), and obtained its phase-coherent timing solution using observations spanning 6 years. M13I has a spin period of 6.37 ms, and an orbital period of 18.23 days. The eccentricity of the binary orbit is 0.064, with a companion mass range of approximately 0.45 to 1.37\,M$_{\odot}$. The orbital properties of M13I are remarkably different from those of the other known pulsars in M13, indicating that this pulsar has undergone a different evolutionary path compared to the rest.
\end{abstract}


\section{Introduction} \label{sec:intro}

Globular Clusters (GCs) are dense, spherical assemblies of millions of stars that are gravitationally bound.
To date, a total of 345 pulsars have been identified across 45 GCs\footnote{\url{https://www3.mpifr-bonn.mpg.de/staff/pfreire/GCpsr.html}}\label{foot:GCpsr}, including 148 isolated and 197 binary pulsars. The majority of these are millisecond pulsars, which are thought to have undergone a "recycling" process involving mass and angular momentum transfer from their companion stars \citep{2017MNRAS.471..857F}.

In recent years, the discovery of pulsars in GCs has been facilitated by the use of large radio telescopes equipped with highly sensitive receivers. 
For example, several pulsars discovered using the Ultra-Wideband Low receiver of the 64-meter Parkes radio telescope (also known as Murriyang) \citep[e.g.][]{2020Dai}.  Over one hundred pulsars in GCs have been discovered with MeerKAT Radio Telescope\footnote{\url{https://trapum.org/discoveries/}} \citep[e.g.][]{2021MNRAS.504.1407R,2022Douglas,2022Vleeschower}. Five-hundred-meter Aperture Spherical radio Telescope \cite[FAST, ][]{2011IJMPD..20..989N}  is actively conducting GC pulsar surveys, leading to the discovery of 61 pulsars in GCs to date. These include several notable systems: one of the first pulsars discovered by FAST in GCs \citep{2020ApJ...892L...6P}, pulsar with the shortest orbital period in GCs \citep{2023Natur.620..961P}, pulsar with the longest spin period in GCs \citep{2024ApJ...974L..23W,2024SCPMA..6769512Z} and  17 pulsars discovered in GC NGC 6517, making it the GC with the highest (or third-highest) number of known pulsars in the FAST sky (or all-sky surveys) \citep[Dai et al., in prep]{2021ApJ...915L..28P,2021RAA....21..143P,2024ApJ...969L...7Y}. The majority of FAST GC discoveries were using the Fast Fourier Transform (FFT) method.

The application of Fast Folding Algorithm (FFA) \citep{1969IEEEP..57..724S} in pulsar search has gradually gained attention. Previously, the use of FFA included the radio signal searching in X-ray pulsar binary systems or isolated X-ray neutron stars \citep{2009ApJ...696..574C,2009ApJ...702..692K}, Processing data from some Rotating RAdio Transient pulsars \citep{2014cosp...40E1880L,2014ARep...58..537L}. Compared with the FFT method, FFA has higher sensitivity in the whole pulsar parameter space, specifically when dealing with ideal data \citep{2020MNRAS.497.4654M}. The superiority of FFA has also been verified in actual observational data. For instance, in the PALFA survey, several pulsars that were overlooked with FFT method but successfully detected through the application of the FFA \citep{2018ApJ...861...44P}. FFA is more sensitive than FFT, but previous FAST GC pulsar survey did not widely employ FFA. Therefore, we decided to search for pulsars in FAST GC data using FFA with the aim of discovering more potential pulsar signals.
The FFA-based software package \textsc{RIPTIDE}\footnote{\url{https://github.com/v-morello/riptide}} can meet the speed requirements for this work. 

In this study, we used the FFA to search pulsars in GCs within the FAST sky. 
We aimed to leverage the sensitivity of FAST and the advantage FFA to search for new pulsars, especially those with long spin period in GCs. Section 2 provides an overview of the observational data and search methodology. Section 3 presents the results of our search. In Section 4, we discuss the implications of our findings, and finally, Section 5 summarizes the key results of this work.

\section{Observations and Search Method} \label{sec:obs}
\subsection{Observation}
The data used in this work were obtained using the 19-beam receiver of FAST \citep{2020Jiang}, as part of the FAST GC pulsar survey. The detailed information of FAST GC survey can be found in \citet{2021ApJ...915L..28P,2024ApJ...974L..23W,2024ApJ...969L...7Y,2024ApJ...972...43L}. Observations were conducted over a frequency range of 1 to 1.5\,GHz. The received electromagnetic signals were processed through channelization and digitization using a Reconfigurable Open Architecture Computing Hardware (ROACH) unit, developed by the Collaboration for Astronomy Signal Processing and Electronics Research (CASPER) group 
(see \url{http://casper.berkeley.edu/}). Following this, the data were packetized and stored in search-mode PSRFITS format. The data had 4096 channels, which results a channel bandwidth of 122\,kHz, and a sampling time of 49.152 $\mu s$. For further details regarding the FAST system, see \citep{2019SCPMA..6259502J}. Table \ref{tab:1} listed some detailed information of the GCs in this work, including the number of observations, data length, dispersion measure (DM) range of each GC and the shortest search period in our work.
\subsubsection{Previous Searches for GC M13}
The GC M13  has been regularly monitored with FAST since 2017, with follow up timing observations conducted approximately once per month. Initial discoveries, including M13F, were reported by \citet{2020ApJ...892...43W},using the PRESTO pulsar search toolkit \textsc{PRESTO}\footnote{\url{https://github.com/scottransom/presto}} \citep{2002AJ....124.1788R}, which also provided timing solutions for M13A–F. Subsequent data were systematically searched every six months using \textsc{PRESTO} following the methodology detailed in \citet{2020ApJ...892...43W}. This ongoing campaign led to the discovery of two additional pulsars, M13G and M13H, with further analysis to be presented in Yin et al. (in prep). In parallel, we processed the data using the FFA pipeline described in this work.

\begin{table*}[h]
\centering
\caption{Some basic information about the GCs searched for in this work.}
\label{tab:1}
\begin{tabular}{llllllllll}
\hline\hline
Cluster     & Obs  & Duration (h) & DM$_{\rm min}$  (pc cm$^{-3}$)  & DM$_{\rm max}$  (pc cm$^{-3}$)  & $P_{min}$ (ms) &  Number of pulsars & Detected with FFA \\ 
M2                  & 19        & 37.5                          & 42.5 & 45.5 & 4   & 10 & 10                      \\ 
M3            &       6      &        15.0         &     25.7 & 27.0      &      4      & 5 & 5                   \\ 
M5             &       14     &          20.5           &    27.0 & 33.0       &       4    & 7 & 7                    \\ 
M10           &      6      &  10.8            &     42.0 & 46.0      &                   4  & 2 & 2  \\ 
M12          &      6             &             11.3           &    40.0 & 45.0       &   4   & 2 & 1   \\ 
M13           &      78             &    96.0            &     27.0 & 34.0      &        3   & 9 & 8 \\ 
M14           & 24                 &    29.6                           & 75.0 & 85.0          &   4  & 5 & 3         \\ 
M15           &   19             &               47.5                &      65.0 & 68.5     &  4   & 15 & 14   \\ 
M53          &   17                &              47.6                 &     22.0 & 28.0      &         4  & 5 & 5  \\ 
M71           &       30           &              49.8                 &    105.0 & 125.0       &        4 & 5 & 5   \\ 
M92          &    23              &               24.5                &     32.5 & 36.5      &          3  & 2 & 2 \\ 
NGC 6517       &    7             &                14.8               &    170.0 & 190.0       & 4   & 21 & 20  \\ 
NGC 6539      &      5          &            5.6                   &      165.0 & 205.0     &       4  & 1 & 1 \\ 
NGC 6712       &      10       &               11.0                &   140.0 & 170.0        &  3   & 1 & 1  \\ 
NGC 6749       &        5       &               7.3                &     180.0 & 205.0      &         3   & 1 & 1  \\ 
NGC 6760     &        1       &      0.8    &      190.0 & 210.0     &       3  & 2 & 2 \\ 
IC 1257      &    1            &                3.0               &    10.0 & 250.0       &         50  & 0 & 0    \\ 
IC 1276    	  &    2              &                3.5               &    10.0 & 400.0       & 3      & 0 & 0    \\ 
M56       &    1            &                2.0               &    10.0 & 360.0       & 5   & 0 & 0   \\ 
Mercer5       &    1         &                0.4               &    10.0 & 700.0       & 20    & 0 & 0 \\ 
NGC 2419     &    2             &                7               &    10.0 & 190.0       & 10     & 0 & 0 \\ 
NGC 5466     	  &    1           &                5.0               &    10.0 & 150.0       & 50  & 0 & 0  \\ 
NGC 6229     &    3             &                2.8               &    10.0 & 120.0       & 4    & 0 & 0 \\ 
NGC 6366     	  &    1             &                2.0               &    10.0 & 230.0       & 4   & 0 & 0  \\ 
Pal10      &    1            &                0.8               &    10.0 & 400.0       & 6   & 0 & 0  \\ 
Pal11       &    2           &                1.0               &    10.0 & 280.0       & 6  & 0 & 0  \\ 
Pal13      &    1         &                0.8               &    10.0 & 100.0       & 6    & 0 & 0 \\ 
Pal14       &    1        &                0.8               &    10.0 & 100.0       & 6    & 0 & 0 \\ 
Pal15       &    1        &                0.8               &    10.0 & 220.0       & 6    & 0 & 0  \\ 
Pfleiderer2       &    1         &                0.8               &    10.0 & 300.0       & 50   & 0 & 0  \\ \hline
\end{tabular}
\begin{flushleft}  
\textbf{Note:} The first column is the name of the GCs. The second and third columns represent the number of observations and the total duration of these observations respectively. The fourth and fifth columns are the minimum and maximum values for de-dispersion during processing. $P_{min}$, represents the shortest search period set during the search process. 
The last two columns show the number of known pulsars in GCs and the number that can be detected by FFA. Among these, the first 16 GCs contain known pulsars, while the last 14 GCs have not reported with known pulsars yet.  
\end{flushleft}
\end{table*}

\subsection{Search Method} \label{sec:search}
The software packages used to process the data are \textsc{PRESTO}\citep{2002AJ....124.1788R} and \textsc{RIPTIDE} (version 0.2.5)\citep{2020MNRAS.497.4654M}. We firstly masked radio frequency interference (RFI) using the {\tt rfifind} routine in the \textsc{PRESTO} for all data set. The de-dispersion strategy were designed as follow: 

\begin{enumerate}
    \item \textbf{GCs with 3 or more known pulsars and DM $< 100\,\rm pc\cdot cm^{-3}$}: use the mean DM ($\overline{ \rm DM}$) of known pulsars as centre DM and 2 $\times$ $\Delta$DM ($\Delta$DM = max. DM $-$ min. DM) as DM range, 
    \item \textbf{GCs with 3 or more known pulsars and DM $> 100\,\rm pc\cdot cm^{-3}$}:use $\overline{\rm DM}$ of known pulsars as centre DM and $\overline{\rm DM} \pm 0.1 \times \overline{\rm DM}$ as DM range,
    \item \textbf{GCs with 2 to 3 known pulsars:} use $\overline{ \rm DM}$ of known pulsars as centre DM and 4,5 or 20 $\rm pc\cdot cm^{-3}$ as DM range according to $\overline{ \rm DM}$,
    \item \textbf{GCs with one known pulsar:} use the DM of the pulsar as centre DM and DM $\pm$ 0.1 $\times$ DM as DM range,
    \item \textbf{GCs without known pulsars}: use DM estimated from YMW16 model \citep{2017ApJ...835...29Y}  as centre value and 10--2$\times$DM as DM range.
\end{enumerate}

According to these strategies, the DM ranges used for each GC are presented in Table \ref{tab:1}. For all cases, we adopted a DM step size of 0.1 $\rm pc\cdot cm^{-3}$ for the de-dispersion process. This choice optimizes computational efficiency while maintaining an acceptable level of smearing. and according to equation \ref{eq:1}, smearing due to an incorrect DM is 89.5 $\mu$s, which is only 3\% of the shortest search period of 3 ms. For GCs with previously unknown DMs, we estimated the values using the Galactic electron density model (YMW16; \citealt{2017ApJ...835...29Y}). This estimation was based on the positional coordinates and distances of the clusters as provided by the Harris catalog \citep{1996AJ....112.1487H}.


\begin{equation}  
t_{\Delta \rm DM} = 4.15 \left( \left( \frac{\nu_{\rm low}}{\mathrm{GHz}} \right)^{-2} - \left( \frac{\nu_{\rm high}}{\mathrm{GHz}} \right)^{-2} \right) \left( \frac{\Delta \rm DM}{\mathrm{pc} \cdot \mathrm{cm}^{-3}} \right) \, \mathrm{ms},
\label{eq:1}
\end{equation} 


Then  {\tt prepsubband} was used together with the rfi mask file to generate a number of de-dispered times series with 512 subbands. 
After de-dispersing the data, the {\tt realfft} algorithm was applied to transform the de-dispersed time series into the frequency domain. Subsequently, the {\tt rednoise} algorithm from \textsc{PRESTO} was employed to mitigate potential red noise contamination in the power spectrum. While the \textsc{RIPTIDE} pipeline incorporates inherent red noise mitigation, our empirical testing revealed that preprocessing the data with the {\tt rednoise} algorithm prior to search implementation yields a substantial reduction in the number of candidates (from 462 to 174 in the 2021 December 9 M10 observation) while maintaining signal fidelity. Notably, the known pulsar M10A exhibited only a marginal Signal-to-Noise Ratio (SNR) variation (14.08 to 13.99) following this additional processing step. To optimize efficiency during candidate screening, we implemented the {\tt rednoise} algorithm for data preprocessing. This red noise suppression procedure was followed by an inverse Fast Fourier Transform (iFFT) operation using {\tt realfft} to generate a whitened time series. For subsequent periodicity searches, we applied the {\tt rffa} command from the \textsc{RIPTIDE} package to analyze the whitened time series in time domain.
Further details about the \textsc{RIPTIDE} package can be found in \citet{2020MNRAS.497.4654M}. The search procedure was initiated in GCs with previously identified pulsars. As an initial test of the entire pipeline, we began by recovering all known signals in these GCs. After redetectining the known pulsar signals, these contributions were subtracted from the whitened time series in frequency domain using the {\tt zapbirds} command in \textsc{PRESTO} to further reduce the number of candidates.  This step is because strong pulsar signals often generate harmonic artifacts at integer multiples of their intrinsic periods, significantly complicating signal identification. A representative case is NGC 6539A, the isolated known pulsar in its host cluster. During a 30-minute FAST observation on 2019 June 20, this source exhibited an exceptionally high SNR of 351 in our FFA analysis. Initial processing yielded 1,161 candidate signals, but application of {\tt zapbirds} reduced this to 637 viable candidates by effectively removing harmonic contamination. After eliminating signals from known pulsars, we performed an inverse FFT to reconstruct the time series in time domain before conducting further signal searches as mentioned above. This process allowed us to perform a subsequent search for any potential new signals.

The computational cost of the FFA search scales approximately as ${p_{min}} ^{-2}$, where $P_{min}$ represents the minimum search period \citep{2020MNRAS.497.4654M}. This inverse-square relationship demonstrates that searches for shorter-period pulsars require substantially greater computational resources. To optimize processing efficiency within our available computational budget, we implemented cluster-specific minimum period thresholds, with these values cataloged in Table \ref{tab:1}.

After completing the search, candidates with a SNR exceeding 7 were selected for further analysis. The {\tt prepfold} command from the \textsc{PRESTO} software suite was then used to fold these candidates. This process generated diagnostic plots that included profiles as a function of time, and DMs. These plots were subsequently examined through visual inspection to determine the signal was whether from a pulsar.
\section{Results} \label{sec:floats}

\subsection{Re-detected of known pulsars and search pulsar in other GCs}

We searched pulsars in 16 GCs which contain known pulsars within the FAST sky coverage using the method describe in Section \ref{sec:search}. During the data analysis, we found that the FFA demonstrated a high detection rate for known pulsars. Data from these GCs containing 93 known pulsars were analyzed. Among these, 39 are isolated pulsars, and 38 can be detected by the FFA. The remaining 54 pulsars are in binary systems, and the FFA detected 49 of them. 
The number of known pulsars and detected pulsars in each GC are listed in Table \ref{tab:1}.
Those non-detected pulsars with FFA are summerized in Table \ref{tab:2}. M12B and M13H exhibit apparent period derivatives exceeding $10^{-11}\ \mathrm{s/s}$ (Yu et al., Yin et al., in prep). For each of M14D, M14E, M15C and NGC 6517V, a known ephemeris was available and we used it to phase-coherently fold the data using \textsc{PRESTO}'s {\tt prepfold}, and they show weak signals with SNR below 7.
These factors contributes to the non-detection of these pulsals using the methods employed in this work.


\begin{table*}[h]
\centering
\caption{The non-detected pulsars using the FFA pipeline. The last column gives potential reason for the non-detection and these information were obtained from \textsc{PRESTO} search.}
\label{tab:2}
\begin{tabular}{llllll}
\hline\hline
Pulsar Name      & Spin period (ms)  & Orbital period (days) & Notes  &   \\ 
M12B & 2.76 & 0.50 & significant acceleration &     \\
M13H  & 11.21  & unknown  & significant acceleration   &      \\
M14D      & 2.89 & 0.74 & significant acceleration &  \\ 
M14E & 2.28   & 0.85   &  significant acceleration  &  \\ 
M15C   & 30.53   & 0.34   &  SNR$<$7  &   \\ 
NGC 6517V & 4.55  & / & SNR$<$5 \\
\hline
\end{tabular}
\end{table*}

Using the method described in Section \ref{sec:search}, we conducted a search of 14 GCs that have not yet been reported to host pulsars. These clusters include IC 1257, IC 1276, M56, Mercer 5, NGC 2419, NGC 5466, NGC 6229, NGC 6366, Pal 10, Pal 11, Pal 13, Pal 14, Pal 15, and Pfleiderer 2. A total of 138448 candidates were identified during the search, however, no new pulsars were discovered.


\subsection{The New pulsar M13I}

\begin{figure*}[h]
\centering 
\includegraphics[width=0.9\linewidth]{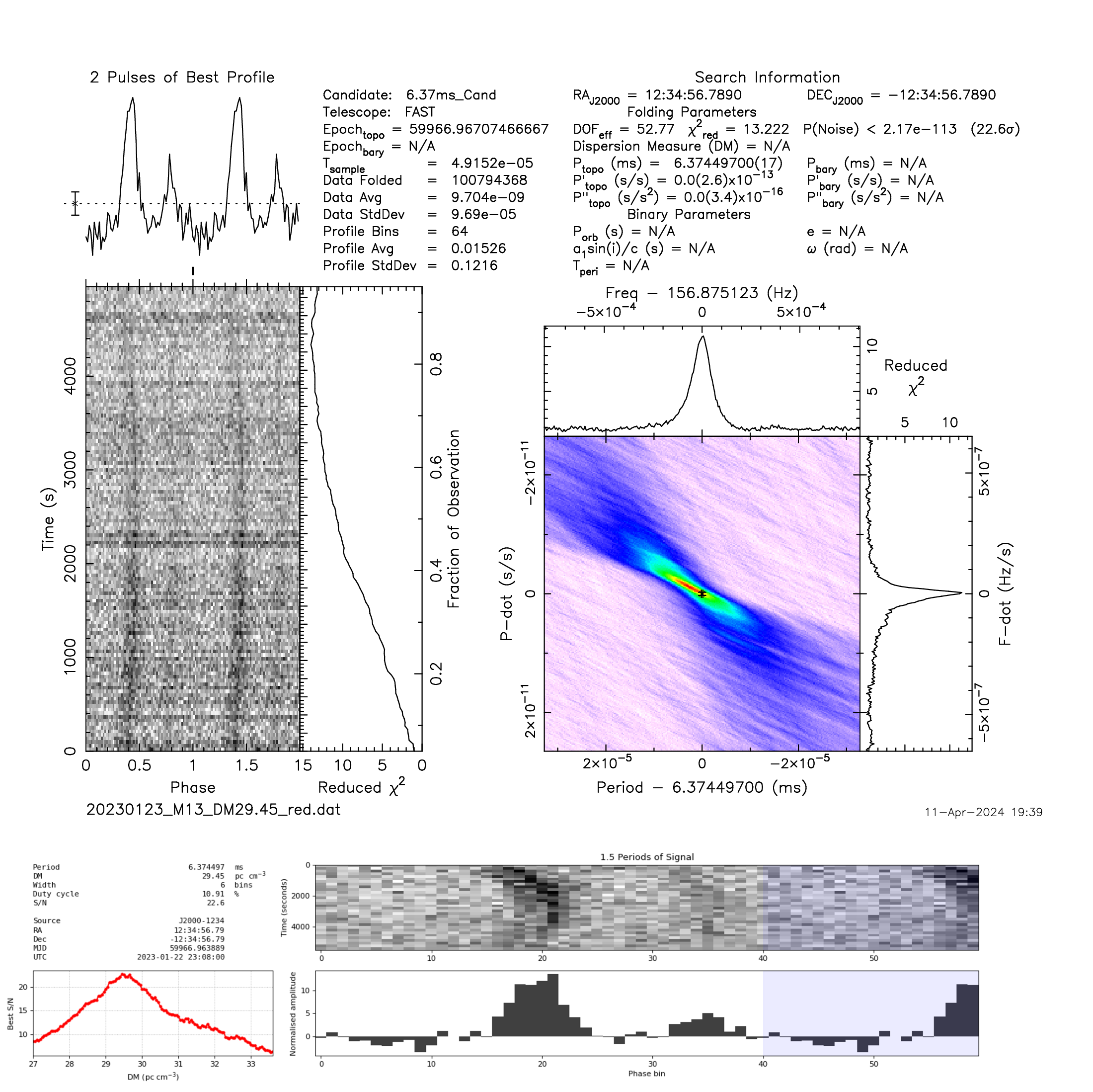}
\caption{The discovery plot of M13I. The observation date is January 23, 2023. The lower part of the plot shows the candidate plot generated by \textsc{RIPTIDE}, while the upper part displays the folded plot of the same signal using \textsc{PRESTO}.}
\label{fig:1}
\end{figure*}

We discovered a new pulsar, designated as M13I (or J1641+3624I), in the GC M13 (Figure \ref{fig:1}). Across 78 observations, M13I was detected 14 times, corresponding to a detection rate of 17.9\%. M13I was initially overlooked in earlier searches due to its weak signal prior to 2023. This pulsar was first identified through our FFA pipeline (see Section \ref{sec:search}) in the January 2023 dataset and subsequently detected via a PRESTO search of the same data. However, archival PRESTO processing of observations up to September 14, 2019, revealed only a single-day detection of M13I, highghting its previously marginal detectability. Figure \ref{fig:4} gives the SNRs of M13I detected using both FFA and FFT respectively.

This pulsar has a spin period of 6.37\,ms and a DM of 29.5 $\rm pc\cdot cm^{-3}$. These characteristics make M13I the pulsar with the third-longest spin period and the second-lowest DM in M13. 
A standard polarisation calibration procedure was performed via the {\sc pac} routine in {\sc psrchive} \citep{Hotan2004}.RM values were calculated using the data observed in January 2023, which is the brightest observation, with the {\sc rmfit} routine in {\sc psrchive} software package.
The profile of M13I displays a prominent main pulse and a distinct inter-pulse, separated by approximately 0.5 of the rotational period (See  Figure \ref{fig:3}). Figure \ref{fig:3} also displayed the linear and circular polarisations of M13I, and the main pulse almost show 100\% linear polarisation. The polarisation angle (PA) showed a flat distribution in upper panel of Figure \ref{fig:3}. The measured rotation measure (RM) of M13I was 20(30) $\text{rad}\cdot\text{m}^{-2}$, which was in agreement with RMs of M13A to F \citep{2023RAA....23j4002W}, but we emphasis that the derived RM value for M13I is not significant.

\begin{figure*}[h]
\centering 
\includegraphics[width=1.0\linewidth]{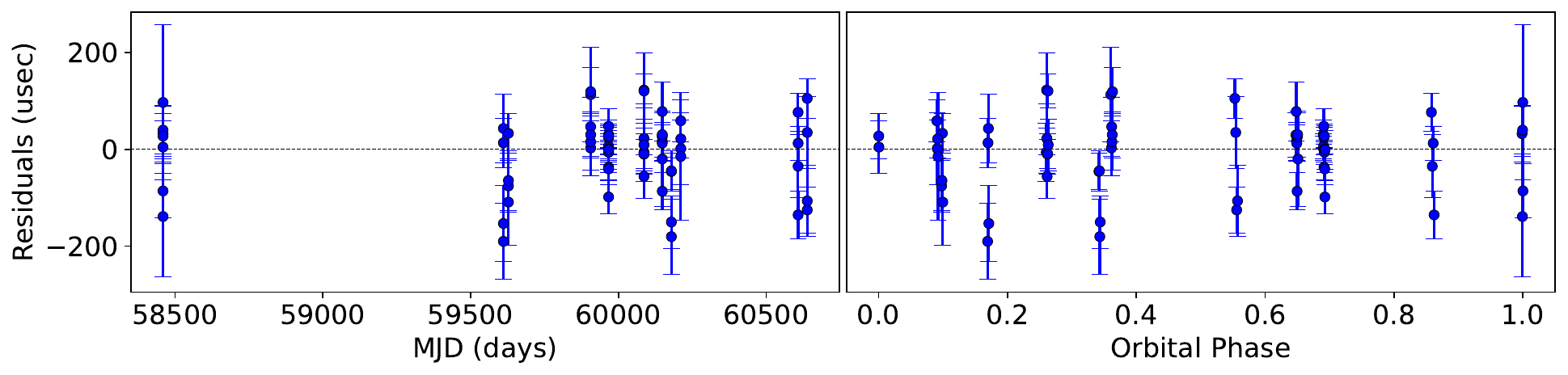}
\caption{The left and right of the plot are the time residuals from the best-fit timing model and the timing residuals as a function of orbital phase for the M13I, respectively.}
\label{fig:2}
\end{figure*}

\begin{figure}[h]
\centering 
\includegraphics[width=1.0\linewidth]{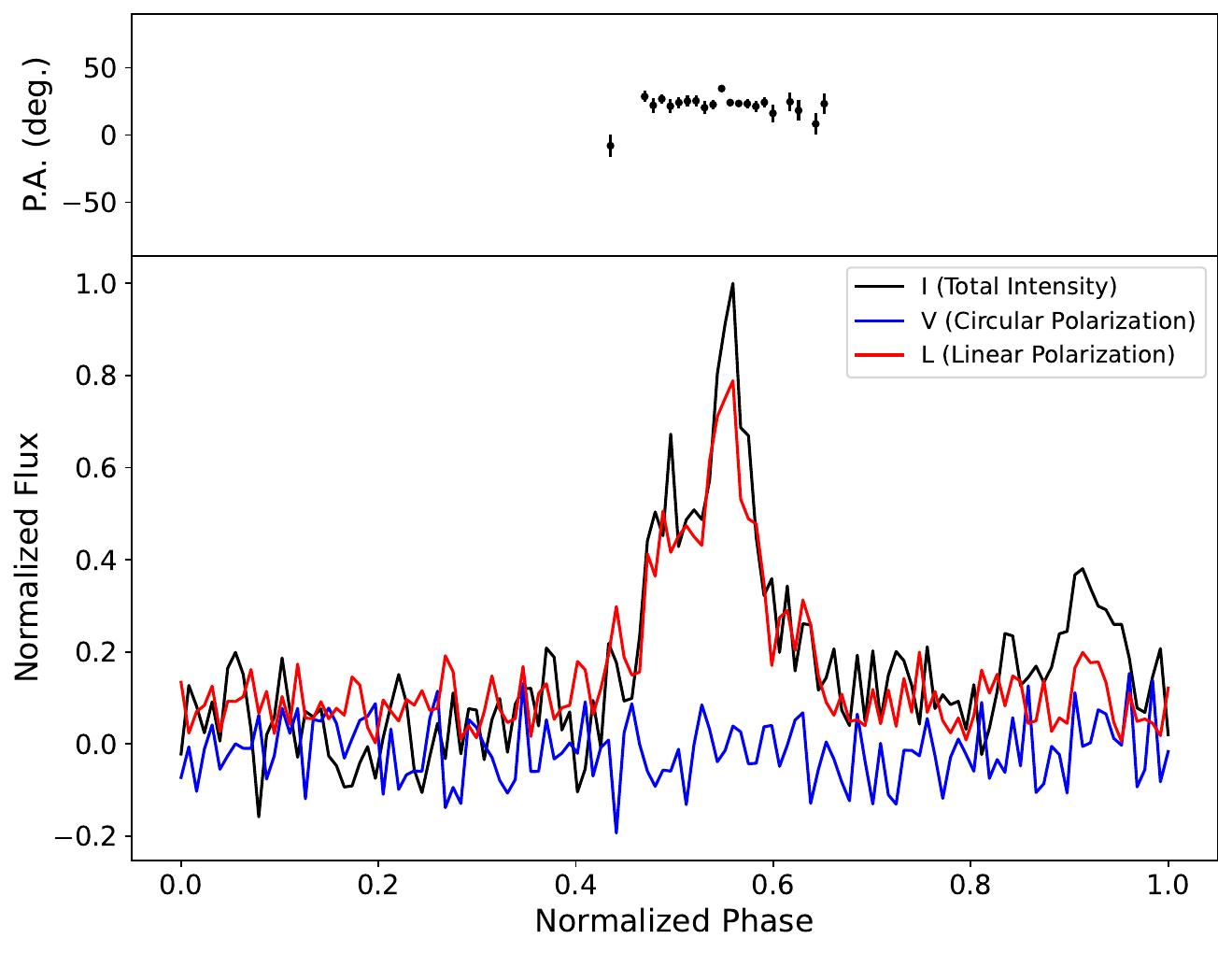}
\caption{The plot shows the polarization calibrated pulse profile of M13I measured at FAST in 2023 - MJD~59966 (the black line represent the total polarization profile, the red line represent the linear polarization profile, and the blue line represent the circular polarization profile).The PA are shown in the upper panel.}
\label{fig:3}
\end{figure}

To determine the orbital parameters of the system, we used the routine {\tt fitorb.py}\footnote{https://github.com/scottransom/presto/blob/master/bin/fitorb.py} from \textsc{PRESTO} to perform an initial fit of the binary parameters. And then we used the command {\tt pygaussfit.py} from \textsc{PRESTO} to fit the detection pulse profile with the highest SNR, obtaining a standard reference profile template for correlation to measuring Times of Arrivals (ToAs). Then the command {\tt get\_TOAs.py} from \textsc{PRESTO} was used to obtain the ToAs. Depending on the strength of the pulsar signal, the number of ToAs obtained per observation ranged from 2 to 16. The measurements from {\tt fitorb.py} were used as initial guesses, and we performed the timing analysis using \textsc{TEMPO}\footnote{\url{http://pulsar.princeton.edu/tempo}} to obtain more precise timing parameter measurements. This process involved several iterations, and we ultimately arrived at a timing solution as shown in Table \ref{tab:3}. The timing residuals are shown in Figure \ref{fig:2}. The DM was measured using ToAs from multifrequency subbands with \textsc{TEMPO}. 



\begin{table}[h]  
\centering
\caption{Timing solution of the new pulsar M13I.\label{tab:3}}
{\footnotesize
\begin{tabular}{ll}  
\hline\hline  
\textbf{Parameter} & \textbf{Value} \\
\hline
Pulsar & M13I \\
\hline
Right ascension, $\alpha$ (J2000) & 16:41:42.015(1) \\
Declination, $\delta$ (J2000) & +36:27:36.522(8) \\
Spin frequency, $f$ (s$^{-1}$) & 156.8583852170(1) \\
First spin frequency derivative, $\dot{f}$ (Hz s$^{-2}$) & $-5.575(1) \times 10^{-14}$ \\
Second spin frequency derivative, $\ddot{f}$ (Hz s$^{-3}$) & $1.3(3) \times 10^{-24}$ \\
Start of timing data (MJD) & 58459.189 \\
End of timing data (MJD) & 60639.194 \\
Dispersion measure, DM (pc cm$^{-3}$) & 29.58(1) \\
Number of ToAs & 68 \\
Residuals RMS ($\mu$s) & 53.31 \\
Binary model & BT \\
Projected semi-major axis, $A1$ (lt-s) & 20.71515(1) \\
Orbital period, $P_b$ (days) & 18.233778(3) \\
Epoch of passage at periastron, $T_0$ (MJD) & 59972.6179(1) \\
Orbital eccentricity, $e$ & 0.064356(3) \\
Longitude of periastron, $\omega$ (deg) & 251.037(2) \\
Minimum companion mass (\,M$_{\odot}$) & 0.4537 \\
Median companion mass (\,M$_{\odot}$) & 0.5406 \\
Maximum companion mass (\,M$_{\odot}$) & 1.3682 \\
\hline\hline
\end{tabular}
}
\begin{flushleft}
    \textbf{Note:} The companion masses are calculated assuming a pulsar mass of 1.4 M$_{\odot}$.
\end{flushleft}
\end{table}

M13I located in a binary system with an orbital period of 18.23 days.
Its eccentricity is 0.064, the median companion mass is about 0.54\,$M_{\odot}$, and the minimum companion mass and maximum companion mass are about 0.45\,$M_{\odot}$ and 1.37\,$M_{\odot}$, respectively.
Previously, the longest orbital period in M13 was for M13F, which is 1.378 days, while the largest eccentricity and companion mass were for M13D, with values of 0.0005 and 0.21\,$M_{\odot}$, respectively. 


\section{Discussion}

\subsection{sensitivity comparison}
Within the sample of 39 known isolated pulsars across these 16 GCs, NGC 6517V represents the sole pulsar not detected through our FFA analysis. This particular pulsar was originally identified via a power spectrum stacking technique applied to all FAST observational data (Dai et al., in prep). We conducted additional searches for NGC 6517V using FAST's standard GC pulsar search pipeline, which employs a FFT-based approach  \citep{2021ApJ...915L..28P}, across the complete NGC 6517 dataset. However, these searches failed to yield any significant detection of the pulsar signal. Among the 54 binary pulsars in our sample, the FFA successfully detected 49 systems. The five undetected pulsars in our FFA search include M12B and M13H, both exhibiting significant orbital effects, as evidenced by their exceptionally high period derivatives ($\dot{\rm P}$ > $10^{-11}\ \mathrm{s/s}$) measured over 1-hour integrations (Yu et al., Yin et al., in prep). These extreme values are dominated by binary accelerations. These accelerations induce a quadratic phase variation in the pulsar signal, described by \(\Phi = \nu t + \frac{1}{2} \dot{\nu} t^{2}\), where the second term arises from the binary acceleration \citep{1991JK}. When folding data using the FFA at a fixed period, this term introduces a quadratic phase drift. A pulsar signal becomes undetectable if the accumulated phase drift exceeds one cycle, i.e., \(\frac{1}{2} \abs{\dot{\nu}} \rm T^{2} \geq 1\), even when folded at the best-fit period. Here T is the total integration time. Substituting \(\nu=\frac{1}{\rm P}\) and \(\dot{\nu}=-\frac{\dot{\rm P}}{\rm P^{2}}\),  the detectability condition becomes \(\abs{\dot{\rm P}} \leq 2 (\frac{\rm P}{\rm T})^{2}\). For a typical integration time T = 1 hour and a pulsar period P = 3\,ms, the maximum detectable period derivative \(\abs{\dot{\rm P}_{max}}\) for detectability is \(1.4 \times 10^{-12}\ \mathrm{s/s}\). However, the measured period derivatives of M12B and M13H exceed \(10^{-11}\ \mathrm{s/s}\), far above the detectability threshold. This quantitatively explains their non-detection via FFA. As reported by \citet{2021ApJ...915L..28P}, pulsars M14D and M14E demonstrate significant orbital acceleration effects, with their signals drifting substantially within 2-hour observations (see Figure 1 in \citealt{2021ApJ...915L..28P}). This rapid drift, combined with their intrinsically weak emission, likely accounts for their absence in our FFA detections. Notably, in our dataset, M15C was detected one time via FFT-based searches with marginal SNR (SNR $\sim$ 7) \citep{2024ApJ...974L..23W}, but it remained undetected in our FFA analysis.

\begin{figure}[h]
\centering 
\includegraphics[width=1.0\linewidth]{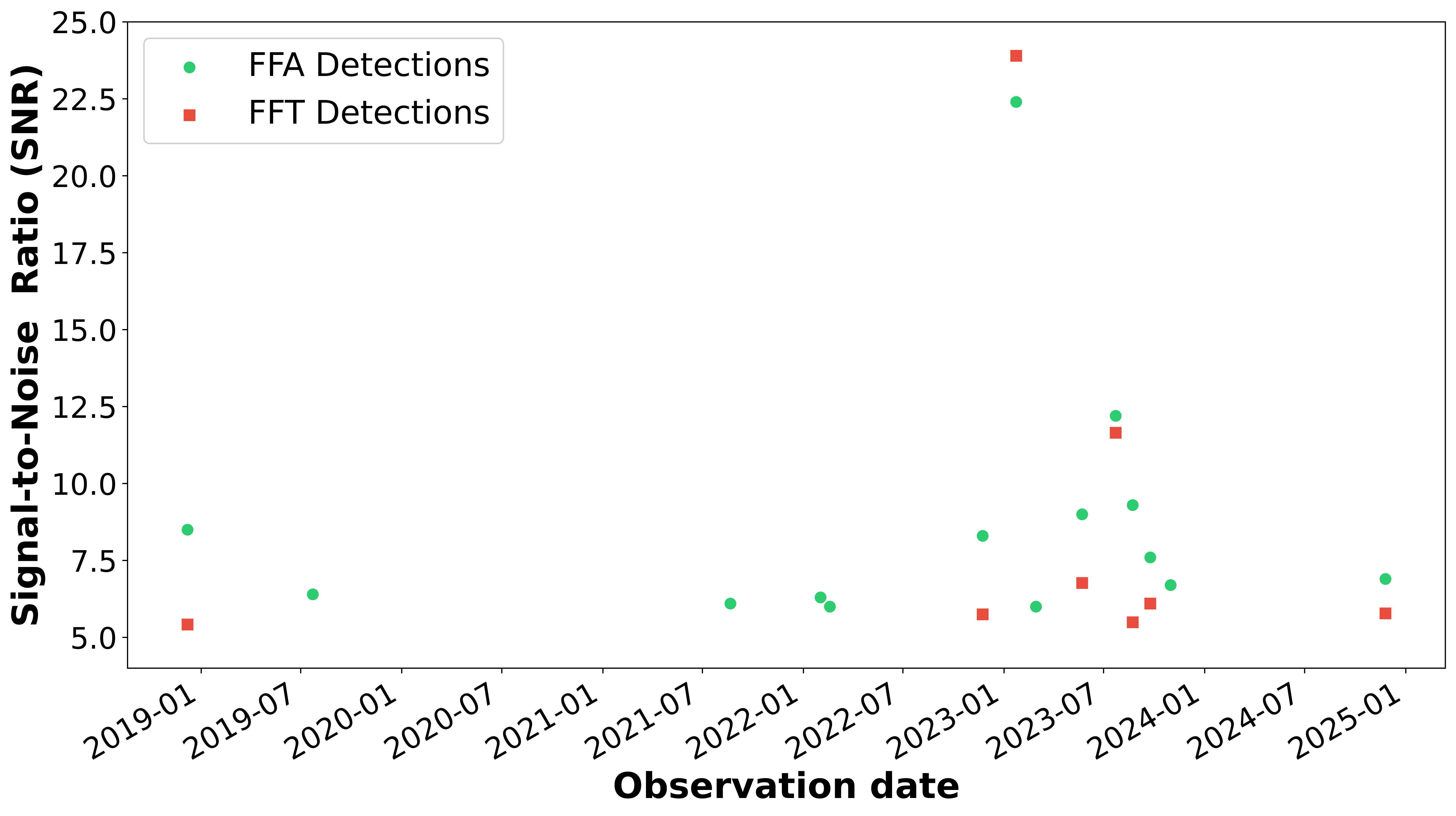}
\caption{A plot of SNR comparison of M13I detections by FFA and FFT. Green circles denote FFA detections (14 instances in total), while red squares represent FFT detections (8 instances). The FFT exhibited a higher SNR than FFA only on the day with the strongest signal, whereas FFA consistently achieved superior SNR in all other observations.}
\label{fig:4}
\end{figure}

For the newly discovered pulsar M13I, as demonstrated in Figure \ref{fig:4}, our analysis revealed a marked difference in detection efficiency between the two search methods: while the FFT technique identified signals in only 8 of 78 observations (10.3\%), the FFA achieved 14 detections (17.9\%), demonstrating a 1.75 times higher detection rate. Notably, with the exception of the brightest epoch (January 2023), where FFT yielded a marginally higher SNR, the FFA method consistently produced superior SNR measurements across all other detections.

Specifically, for signals with spin periods shorter than the shortest search period, our sensitivity is significantly reduced. We can only detect sub-harmonic signals with periods longer than the shortest search period.These sub-harmonics are weaker than the intrinsic signals, which may make them falling below the SNR threshold, missed in the search.

\subsection{Speed comparison}

\begin{figure}[h]
\centering 
\includegraphics[width=1.0\linewidth]{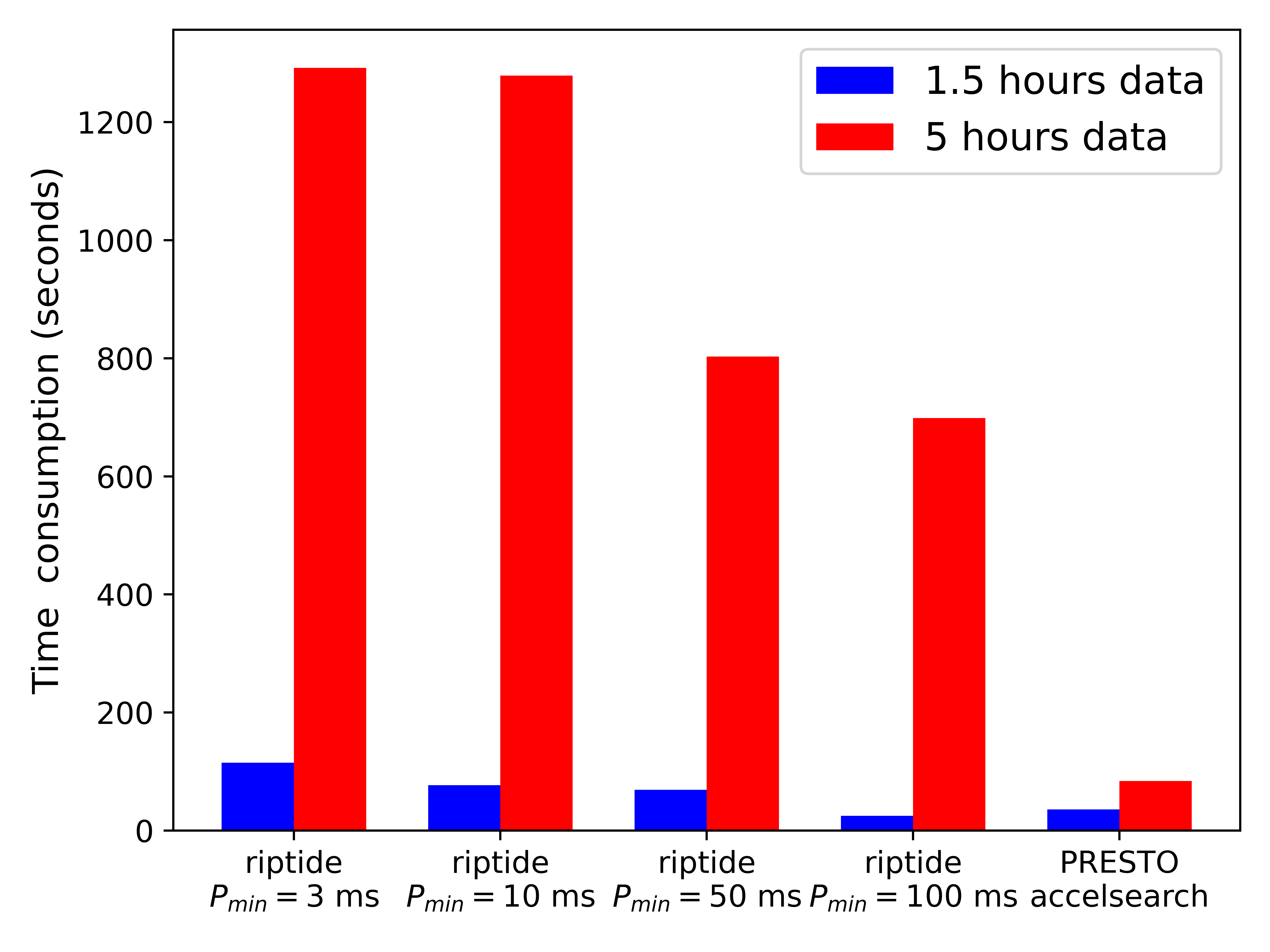}
\caption{A plot showing the time taken for FFA and {\tt accelsearch} with different data lengths. 
The values 3 ms, 10 ms, 50 ms, and 100 ms represent the minimum search periods set for the FFA searches, and {\tt accelsearch} refers to the search conducted using \textsc{PRESTO}.}
\label{fig:5}
\end{figure}

We conducted a test of the processing speed of the search command ({\tt rffa}) in \textsc{RIPTIDE} and the frequency-domain acceleration search ({\tt accelsearch}) in \textsc{PRESTO} using FAST observational data with integration times of 1.5 and 5 hours. For {\tt rffa}, we set the maximum search period (P$_{max}$) to 100 seconds (s), ensuring coverage of all known pulsars in GCs. The minimum search periods (P$_{min}$) were tested at 3\,ms, 10\,ms, 50\,ms, and 100\,ms,respectively. To maintain consistent time resolution across duty cycles, the search range was divided into three segments for each P$_{min}$ configuration. The corresponding parameters—including period bounds (period\_{min}, period\_{max}), binning constraints (bin\_{min}, bin\_{max}) and SNR thresholds—are detailed in Table \ref{tab:seg}. Additionally, we configured the {\tt rffa} algorithm to suppress candidate file generation during processing. For {\tt accelsearch}, we configured   z$_{max}$ = 0 to target isolated pulsars, summing 32 harmonics. The computational time for both algorithms across different P$_{min}$ values is presented in Figure \ref{fig:5}. The results demonstrate that  P$_{min}$ is the dominant factor in efficiency of {\tt rffa}, with the 5-hour dataset requiring significantly more processing time than the 1.5-hour dataset for both methods. Notably, {\tt rffa} generally exhibited longer execution times than {\tt accelsearch}, except for the 1.5-hour dataset at P$_{min}$ = 100\,ms.
Due to limitations in our computational resources, our search was limited to signals with periods $\geq$ 3\,ms, in spite there are several GC pulsars with shorter spin periods.  Future reductions in computational costs could enable the inclusion of signals in the 1--2\,ms range, and potentially sub-millisecond periods, within our FFA pipeline.

\begin{table}[htbp]
    \centering
    \caption{Searching Parameters with {\tt rffa} in three segments.}
{\footnotesize   
    \begin{tabular}{cccccc}
        \toprule
         & period\_min & period\_max & bins\_min & bins\_max & SNR\\
        \midrule
        seg 1 & $P_{min}$ & 0.5 s & 40 & 100 & 7 \\
        seg 2 & 0.5 s & 2.0 s & 300 & 520 & 7 \\
        seg 3 & 2.0 s & 100.0 s & 960 & 1040 & 7 \\
        \bottomrule
  \end{tabular}
  }
  \label{tab:seg}
\end{table}

\subsection{Properties of the pulsar M13I}
We used the radiometer equation , Equation \ref{eq:4}, (\citealt{lorimer2005handbook}) to estimate the mean flux density of M13I.
\begin{equation}
    S_{\mathrm mean}=\frac{\sigma \beta T_{\mathrm sys} }{G\sqrt{n_{\rm p}t_{\mathrm obs}\Delta f } }\sqrt{\frac{W_{\mathrm obs}}{P_{\mathrm spin}-W_{\mathrm obs}} },
\label{eq:4}
\end{equation}
where $\sigma$ is the SNR of the candidate; $\beta$ is the sampling efficiency and the value is 1 for our 8-bit recording system; the system temperature $T_{\rm sys}$ is $20 {\ \rm K}$, the antenna gain ($G$) is $16\,{\rm K\,Jy}^{-1}$ \citep{2019SCPMA..6259502J}; the number of polarizations ($n_{\rm p}$) is 2; the $t_{\rm obs}$ is the integration time in units of s, and the data length range containing M13I signal is 3600 s to 8775 s; the $\Delta f$ is the bandwidth, 400\,MHz; $P_{\rm spin}$ is the pulsar period and $W_{\rm obs}$ is the pulse width.
The flux density range for M13I spans from 1.3 $\mu {\rm Jy}$ to 5.2 $\mu {\rm Jy}$, with a median value of 1.8 $\mu {\rm Jy}$.


In GCs, the environment near the center is more complex and dense. Binaries closer to the center may experience more frequent and intense gravitational perturbations from other stars, which may lead to an increase in eccentricity \citep{1996MNRAS.282.1064H}. \citet{2020ApJ...892...43W} demonstrated an inverse correlation between binary system eccentricity and projected angular distance from the cluster center for pulsars B, D, E, and F in M13. With a second smallest projected angular distance of 11.7\,arcsecond from the cluster center and its notably higher eccentricity relative to other binary pulsars in M13, M13I contradicts the trend reported by \citet{2020ApJ...892...43W}.  Furthermore, M13I exhibits the lowest DM among the seven known pulsars in M13 (A--F and I). This DM characteristic suggests that M13I is likely positioned in the nearer hemisphere of the cluster along our line of sight.

\begin{figure}[h]
\centering 
\includegraphics[width=1.0\linewidth]{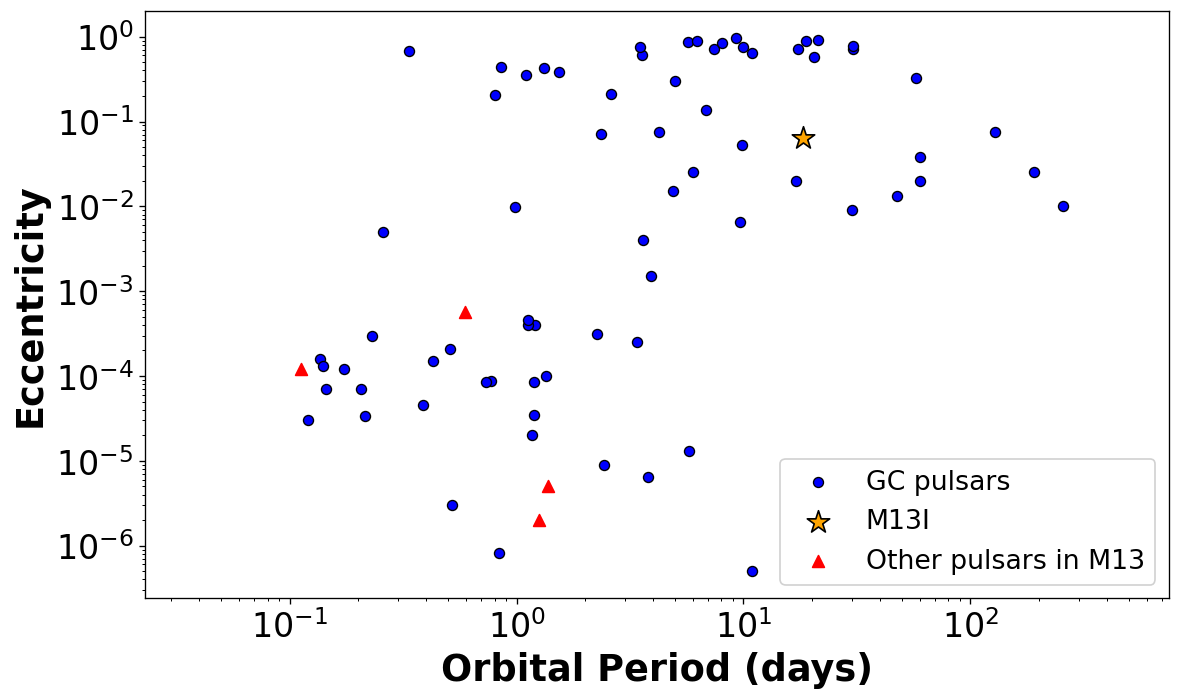}
\caption{Relation between orbital period and eccentricity of pulsars in GCs. Blue dots are binary pulsars in GCs (except M13), red triangles are privious known binary pulsars in M13 (B,D,E and F), and the star represents M13I. Data from website \ref{foot:GCpsr}.}
\label{fig:6}
\end{figure}

Among the 7 pulsars in M13, M13I exhibits the longest orbital period (18.23 days) and highest eccentricity (e = 0.064). In binary systems, tidal interactions between the neutron star and its companion typically lead to orbital circularization over time, as energy dissipation reduces the system's eccentricity \citep{1992RSPTA.341...39P}. However, the large eccentricity of M13I suggests a deviation from this standard evolutionary pathway. To investigate this anomaly, we examined the orbital period--eccentricity distribution of all known GC binary pulsars (Figure \ref{fig:6}). The population shows a bimodal distribution, with M13I occupying the upper-right region (long-period, high-eccentricity systems) while the other M13 binaries cluster in the lower-left region (short-period, low-eccentricity systems). This distinct segregation implies different evolutionary histories between these two groups. Given M13I's exceptionally low flux density (median = 1.8\,$\mu$Jy), its evolutionary analysis presents significant observational challenges. To advance our understanding, may require comparative population studies of pulsars in GCs and detailed dynamical modeling of gravitational interactions within M13's core environment.

\section{Conclusion} \label{sec:cite}
We conducted a search for pulsars in 16 GCs using FAST observational data, employing a FFA-based pipeline. Our pipeline successful detected 38 out of 39 known isolated pulsars, and 49 out of 54 known binary puslars. Six known pulsars were not detected in our search: M12B, M13H, M14D, M14E, M15C, and NGC6517V.

This search led to the discovery of M13I, a new 6.37-ms pulsar residing in a binary system. Comparative analysis revealed that the FFA method achieved a higher detection rate (17.9\%) compared to traditional FFT techniques (10.3\%) in our dataset. These results show that the FFA demonstrates superior sensitivity to weakly accelerated signals.


We derived a coherent timing solution for M13I using a span of 6-year dataset, revealing its orbital parameters: a 18.2-day orbital period and a significant eccentricity of 0.06. The companion star has a median mass of 0.54\,M$_{\odot}$, consistent with a helium white dwarf (He WD) progenitor. The system’s unusually high eccentricity suggests it may have undergone a distinct evolutionary pathway compared to other binary pulsars in M13.


\begin{acknowledgments}
We sincerely appreciate the constructive and insightful comments provided by the anonymous reviewer.
This work is supported by the National SKA Program of China No. 2020SKA0120100,  National Nature Science Foundation of China (NSFC) under Grant No. 12003047 and 12173053. LW gratefully acknowledges support from the Boya postdoctoral fellowship of Peking University, the China Postdoctoral Science Foundation fellowship 2022M720267.
LQ was supported by the Youth Innovation Promotion Association of CAS (id.~2018075, Y2022027), and the CAS "Light of West China" Program.
This work is supported by the National Natural Science Foundation of China under Grand No. 11703047, 11773041, U2031119, and 12173052. ZP is supported by the CAS “Light of West China” Program and the Youth Innovation Promotion Association of the Chinese Academy of Sciences (ID 2023064), National Key R$\&$D Program of China, No. 2022YFC2205202.
LZ has been supported by the National Natural Science Foundation of China (NSFC) under Grant Nos. 12373032, as well as the Science and Technology Program of Guizhou Province under project ZDSYS[2023]003.
R.P.E. is supported by the Chinese Academy of Sciences President's International Fellowship Initiative, grant No. 2021FSM0004.
This work is supported by 100101 Key Laboratory of Radio Astronomy and Technology (Chinese Academy of Science).
This work made use of the data from FAST (Five-hundred-meter Aperture Spherical radio Telescope) (https://cstr.cn/31116.02.FAST). FAST is a Chinese national mega-science facility, operated by National Astronomical Observatories, Chinese Academy of Sciences.

\end{acknowledgments}

\bibliography{file}{}

\begin{thebibliography}{}
\expandafter\ifx\csname natexlab\endcsname\relax\def\natexlab#1{#1}\fi
\providecommand{\url}[1]{\href{#1}{#1}}
\providecommand{\dodoi}[1]{doi:~\href{http://doi.org/#1}{\nolinkurl{#1}}}
\providecommand{\doeprint}[1]{\href{http://ascl.net/#1}{\nolinkurl{http://ascl.net/#1}}}
\providecommand{\doarXiv}[1]{\href{https://arxiv.org/abs/#1}{\nolinkurl{https://arxiv.org/abs/#1}}}

\bibitem[{{Crawford} {et~al.}(2009){Crawford}, {Lorimer}, {Devour}, {Takacs},
  \& {Kondratiev}}]{2009ApJ...696..574C}
{Crawford}, F., {Lorimer}, D.~R., {Devour}, B.~M., {Takacs}, B.~P., \&
  {Kondratiev}, V.~I. 2009, \apj, 696, 574, \dodoi{10.1088/0004-637X/696/1/574}

\bibitem[{{Dai} {et~al.}(2020){Dai}, {Johnston}, {Kerr}, {Camilo}, {Cameron},
  {Toomey}, \& {Kumamoto}}]{2020Dai}
{Dai}, S., {Johnston}, S., {Kerr}, M., {et~al.} 2020, \apjl, 888, L18,
  \dodoi{10.3847/2041-8213/ab621a}

\bibitem[{{Douglas} {et~al.}(2022){Douglas}, {Padmanabh}, {Ransom}, {Ridolfi},
  {Freire}, {Krishnan}, {Barr}, {Pallanca}, {Cadelano}, {Possenti}, {Stairs},
  {Hessels}, {DeCesar}, {Lynch}, {Bailes}, {Burgay}, {Champion}, {Karuppusamy},
  {Kramer}, {Stappers}, \& {Vleeschower}}]{2022Douglas}
{Douglas}, A., {Padmanabh}, P.~V., {Ransom}, S.~M., {et~al.} 2022, \apj, 927,
  126, \dodoi{10.3847/1538-4357/ac4744}

\bibitem[{{Freire} {et~al.}(2017){Freire}, {Ridolfi}, {Kramer}, {Jordan},
  {Manchester}, {Torne}, {Sarkissian}, {Heinke}, {D'Amico}, {Camilo},
  {Lorimer}, \& {Lyne}}]{2017MNRAS.471..857F}
{Freire}, P.~C.~C., {Ridolfi}, A., {Kramer}, M., {et~al.} 2017, \mnras, 471,
  857, \dodoi{10.1093/mnras/stx1533}

\bibitem[{{Harris}(1996)}]{1996AJ....112.1487H}
{Harris}, W.~E. 1996, \aj, 112, 1487, \dodoi{10.1086/118116}

\bibitem[{{Heggie} \& {Rasio}(1996)}]{1996MNRAS.282.1064H}
{Heggie}, D.~C., \& {Rasio}, F.~A. 1996, \mnras, 282, 1064,
  \dodoi{10.1093/mnras/282.3.1064}

\bibitem[{Hotan {et~al.}(2004)Hotan, van Straten, \& Manchester}]{Hotan2004}
Hotan, A.~W., van Straten, W., \& Manchester, R.~N. 2004, Publications of the
  Astronomical Society of Australia, 21, 302–309, \dodoi{10.1071/as04022}

\bibitem[{{Jiang} {et~al.}(2019){Jiang}, {Yue}, {Gan}, {Yao}, {Li}, {Pan},
  {Sun}, {Yu}, {Liu}, {Tang}, {Qian}, {Lu}, {Yan}, {Peng}, {Zhang}, {Wang},
  {Li}, {Li}, \& {FAST Collaboration}}]{2019SCPMA..6259502J}
{Jiang}, P., {Yue}, Y., {Gan}, H., {et~al.} 2019, Science China Physics,
  Mechanics, and Astronomy, 62, 959502, \dodoi{10.1007/s11433-018-9376-1}

\bibitem[{{Jiang} {et~al.}(2020){Jiang}, {Tang}, {Hou}, {Liu}, {Kr{\v{c}}o},
  {Qian}, {Sun}, {Ching}, {Liu}, {Duan}, {Yue}, {Gan}, {Yao}, {Li}, {Pan},
  {Yu}, {Liu}, {Li}, {Peng}, {Yan}, \& {FAST Collaboration}}]{2020Jiang}
{Jiang}, P., {Tang}, N.-Y., {Hou}, L.-G., {et~al.} 2020, Research in Astronomy
  and Astrophysics, 20, 064, \dodoi{10.1088/1674-4527/20/5/64}

\bibitem[{{Johnston} \& {Kulkarni}(1991)}]{1991JK}
{Johnston}, H.~M., \& {Kulkarni}, S.~R. 1991, \apj, 368, 504,
  \dodoi{10.1086/169715}

\bibitem[{{Kondratiev} {et~al.}(2009){Kondratiev}, {McLaughlin}, {Lorimer},
  {Burgay}, {Possenti}, {Turolla}, {Popov}, \& {Zane}}]{2009ApJ...702..692K}
{Kondratiev}, V.~I., {McLaughlin}, M.~A., {Lorimer}, D.~R., {et~al.} 2009,
  \apj, 702, 692, \dodoi{10.1088/0004-637X/702/1/692}

\bibitem[{{Li} {et~al.}(2024){Li}, {Zhang}, {Yao}, {Yin}, {Eatough}, {Li},
  {Li}, {Lian}, {Pan}, {Dai}, {Li}, {Zhang}, {Su}, {Wu}, {Liu}, {Liu}, {Wang},
  {Qian}, \& {Pan}}]{2024ApJ...972...43L}
{Li}, B., {Zhang}, L.-y., {Yao}, J., {et~al.} 2024, \apj, 972, 43,
  \dodoi{10.3847/1538-4357/ad5a82}

\bibitem[{Lorimer \& Kramer(2005)}]{lorimer2005handbook}
Lorimer, D.~R., \& Kramer, M. 2005, Handbook of pulsar astronomy, Vol.~4
  (Cambridge university press)

\bibitem[{{Losovsky} \& {Dmitry Dumsky}(2014)}]{2014cosp...40E1880L}
{Losovsky}, B., \& {Dmitry Dumsky}, M. 2014, in 40th COSPAR Scientific
  Assembly, Vol.~40, E1.15--32--14

\bibitem[{{Losovsky} \& {Dumsky}(2014)}]{2014ARep...58..537L}
{Losovsky}, B.~Y., \& {Dumsky}, D.~V. 2014, Astronomy Reports, 58, 537,
  \dodoi{10.1134/S1063772914080034}

\bibitem[{{Morello} {et~al.}(2020){Morello}, {Barr}, {Stappers}, {Keane}, \&
  {Lyne}}]{2020MNRAS.497.4654M}
{Morello}, V., {Barr}, E.~D., {Stappers}, B.~W., {Keane}, E.~F., \& {Lyne},
  A.~G. 2020, \mnras, 497, 4654, \dodoi{10.1093/mnras/staa2291}

\bibitem[{{Nan} {et~al.}(2011){Nan}, {Li}, {Jin}, {Wang}, {Zhu}, {Zhu},
  {Zhang}, {Yue}, \& {Qian}}]{2011IJMPD..20..989N}
{Nan}, R., {Li}, D., {Jin}, C., {et~al.} 2011, International Journal of Modern
  Physics D, 20, 989, \dodoi{10.1142/S0218271811019335}

\bibitem[{{Pan} {et~al.}(2020){Pan}, {Ransom}, {Lorimer}, {Fiore}, {Qian},
  {Wang}, {Stappers}, {Hobbs}, {Zhu}, {Yue}, {Wang}, {Lu}, {Liu}, {Peng},
  {Zhang}, \& {Li}}]{2020ApJ...892L...6P}
{Pan}, Z., {Ransom}, S.~M., {Lorimer}, D.~R., {et~al.} 2020, \apjl, 892, L6,
  \dodoi{10.3847/2041-8213/ab799d}

\bibitem[{{Pan} {et~al.}(2021{\natexlab{a}}){Pan}, {Qian}, {Ma}, {Liu}, {Wang},
  {Luo}, {Yan}, {Ransom}, {Lorimer}, {Li}, \& {Jiang}}]{2021ApJ...915L..28P}
{Pan}, Z., {Qian}, L., {Ma}, X., {et~al.} 2021{\natexlab{a}}, \apjl, 915, L28,
  \dodoi{10.3847/2041-8213/ac0bbd}

\bibitem[{{Pan} {et~al.}(2021{\natexlab{b}}){Pan}, {Ma}, {Qian}, {Wang}, {Yan},
  {Luo}, {Ransom}, {Lorimer}, \& {Jiang}}]{2021RAA....21..143P}
{Pan}, Z., {Ma}, X.-Y., {Qian}, L., {et~al.} 2021{\natexlab{b}}, Research in
  Astronomy and Astrophysics, 21, 143, \dodoi{10.1088/1674-4527/21/6/143}

\bibitem[{{Pan} {et~al.}(2023){Pan}, {Lu}, {Jiang}, {Han}, {Chen}, {Han},
  {Liu}, {Qian}, {Xu}, {Zhang}, {Luo}, {Yan}, {Yang}, {Zhou}, {Wang}, {Wang},
  {Li}, \& {Zhu}}]{2023Natur.620..961P}
{Pan}, Z., {Lu}, J.~G., {Jiang}, P., {et~al.} 2023, \nat, 620, 961,
  \dodoi{10.1038/s41586-023-06308-w}

\bibitem[{{Parent} {et~al.}(2018){Parent}, {Kaspi}, {Ransom}, {Krasteva},
  {Patel}, {Scholz}, {Brazier}, {McLaughlin}, {Boyce}, {Zhu}, {Pleunis},
  {Allen}, {Bogdanov}, {Caballero}, {Camilo}, {Camuccio}, {Chatterjee},
  {Cordes}, {Crawford}, {Deneva}, {Ferdman}, {Freire}, {Hessels}, {Jenet},
  {Knispel}, {Lazarus}, {van Leeuwen}, {Lyne}, {Lynch}, {Seymour}, {Siemens},
  {Stairs}, {Stovall}, \& {Swiggum}}]{2018ApJ...861...44P}
{Parent}, E., {Kaspi}, V.~M., {Ransom}, S.~M., {et~al.} 2018, \apj, 861, 44,
  \dodoi{10.3847/1538-4357/aac5f0}

\bibitem[{{Phinney}(1992)}]{1992RSPTA.341...39P}
{Phinney}, E.~S. 1992, Philosophical Transactions of the Royal Society of
  London Series A, 341, 39, \dodoi{10.1098/rsta.1992.0084}

\bibitem[{{Ransom} {et~al.}(2002){Ransom}, {Eikenberry}, \&
  {Middleditch}}]{2002AJ....124.1788R}
{Ransom}, S.~M., {Eikenberry}, S.~S., \& {Middleditch}, J. 2002, \aj, 124,
  1788, \dodoi{10.1086/342285}

\bibitem[{{Ridolfi} {et~al.}(2021){Ridolfi}, {Gautam}, {Freire}, {Ransom},
  {Buchner}, {Possenti}, {Venkatraman Krishnan}, {Bailes}, {Kramer},
  {Stappers}, {Abbate}, {Barr}, {Burgay}, {Camilo}, {Corongiu}, {Jameson},
  {Padmanabh}, {Vleeschower}, {Champion}, {Chen}, {Geyer}, {Karastergiou},
  {Karuppusamy}, {Parthasarathy}, {Reardon}, {Serylak}, {Shannon}, \&
  {Spiewak}}]{2021MNRAS.504.1407R}
{Ridolfi}, A., {Gautam}, T., {Freire}, P.~C.~C., {et~al.} 2021, \mnras, 504,
  1407, \dodoi{10.1093/mnras/stab790}

\bibitem[{{Staelin}(1969)}]{1969IEEEP..57..724S}
{Staelin}, D.~H. 1969, IEEE Proceedings, 57, 724,
  \dodoi{10.1109/PROC.1969.7051}

\bibitem[{{Vleeschower} {et~al.}(2022){Vleeschower}, {Stappers}, {Bailes},
  {Barr}, {Kramer}, {Ransom}, {Ridolfi}, {Venkatraman Krishnan}, {Possenti},
  {Keith}, {Burgay}, {Freire}, {Spiewak}, {Champion}, {Bezuidenhout},
  {Ni{\c{t}}u}, {Chen}, {Parthasarathy}, {DeCesar}, {Buchner}, {Stairs}, \&
  {Hessels}}]{2022Vleeschower}
{Vleeschower}, L., {Stappers}, B.~W., {Bailes}, M., {et~al.} 2022, \mnras, 513,
  1386, \dodoi{10.1093/mnras/stac921}

\bibitem[{{Wang} {et~al.}(2020){Wang}, {Peng}, {Stappers}, {Liu}, {Keith},
  {Lyne}, {Lu}, {Yu}, {Kou}, {Yan}, {Jiang}, {Jin}, {Li}, {Li}, {Qian}, {Wang},
  {Yue}, {Zhang}, {Zhang}, {Zhu}, \& {FAST
  Collaboration}}]{2020ApJ...892...43W}
{Wang}, L., {Peng}, B., {Stappers}, B.~W., {et~al.} 2020, \apj, 892, 43,
  \dodoi{10.3847/1538-4357/ab76cc}

\bibitem[{{Wang} {et~al.}(2023){Wang}, {Han}, {Xu}, {Wang}, {Yan}, {Jing},
  {Su}, {Zhou}, \& {Wang}}]{2023RAA....23j4002W}
{Wang}, P.~F., {Han}, J.~L., {Xu}, J., {et~al.} 2023, Research in Astronomy and
  Astrophysics, 23, 104002, \dodoi{10.1088/1674-4527/acea1f}

\bibitem[{{Wu} {et~al.}(2024){Wu}, {Pan}, {Qian}, {Ransom}, {Eatough}, {Wang},
  {Freire}, {Liu}, {Yan}, {Luo}, {Zhang}, {Li}, {Yin}, {Li}, {Li}, {Dai}, {Li},
  {Zhang}, {Liu}, \& {Pan}}]{2024ApJ...974L..23W}
{Wu}, Y., {Pan}, Z., {Qian}, L., {et~al.} 2024, \apjl, 974, L23,
  \dodoi{10.3847/2041-8213/ad7b9e}

\bibitem[{{Yao} {et~al.}(2017){Yao}, {Manchester}, \&
  {Wang}}]{2017ApJ...835...29Y}
{Yao}, J.~M., {Manchester}, R.~N., \& {Wang}, N. 2017, \apj, 835, 29,
  \dodoi{10.3847/1538-4357/835/1/29}

\bibitem[{{Yin} {et~al.}(2024){Yin}, {Zhang}, {Qian}, {Eatough}, {Li},
  {Lorimer}, {Dai}, {Li}, {Zhang}, {Li}, {Su}, {Wu}, {Pan}, {Lian}, {Liu},
  {Yan}, \& {Pan}}]{2024ApJ...969L...7Y}
{Yin}, D., {Zhang}, L.-y., {Qian}, L., {et~al.} 2024, \apjl, 969, L7,
  \dodoi{10.3847/2041-8213/ad534e}

\bibitem[{{Zhou} {et~al.}(2024){Zhou}, {Wang}, {Li}, {Fang}, {Miao}, {Freire},
  {Zhang}, {Zhang}, {Chen}, {Feng}, {Xiao}, {Xie}, {Zhang}, {Jin}, {Wang},
  {Ke}, {Guo}, {Zhao}, {Niu}, {Zhu}, {Xue}, {Wang}, {Wu}, {Gan}, {Sun}, {Wang},
  {Zhang}, {Zhang}, {Cao}, \& {Lu}}]{2024SCPMA..6769512Z}
{Zhou}, D., {Wang}, P., {Li}, D., {et~al.} 2024, Science China Physics,
  Mechanics, and Astronomy, 67, 269512, \dodoi{10.1007/s11433-023-2362-x}

\end{thebibliography}
\bibliographystyle{aasjournal}
\end{document}